\documentclass[a4paper,12pt]{article}
\usepackage[dvips]{graphicx}
\DeclareGraphicsExtensions{.pdf,.jpg}

\usepackage{amsfonts,amsmath,amsthm,amssymb}
\usepackage{mathrsfs}
\usepackage[onehalfspacing]{setspace}
\author{L. Ort\'{i}z\footnote{leonardo.ortiz@uniofyorkspace.net}}

\title{No superradiance for the scalar field in the BTZ black hole with reflexive boundary conditions}

\begin{document}

\maketitle

\begin{center}



Department of Mathematics\\
The University of York\\
York YO10 5DD, U. K.\\\vspace{0.4cm}

\normalsize{\textbf{Abstract}}\\\end{center} \small{We show that
there is no superradiance in the rotating BTZ black hole for
vanishing boundary conditions at infinity for the real scalar
field.}

\section{Introduction}

The superradiance phenomenon takes place when a wave is partially
reflected and partially transmitted. It can be roughly defined by
saying that: the reflected part has a bigger amplitude than the
incident one. This phenomenon can be possible because energy is
being taken from a source which might be for example a
electrostatic field \cite{cMan87}. It turns out that this
phenomenon also occurs on black hole geometries, particularly on
the Kerr metric. However, in this metric the real scalar field
present superradiance whereas a fermionic field not, see for
example \cite{wUnruh74}, \cite{chLee77}. The BTZ metric has some
similarities with the Kerr metric. Hence it seems natural to
investigate what happen in the BTZ metric regarding superradiance.
Apart from the interest in its own right, superradiance is closely
related to quantum effects on black holes, for instance, pair
particle creation.

Naively, we could expect that superradiance must occur on the BTZ
metric for the scalar field whereas must be absent for the Dirac
field. See for example \cite{sCar95}, where it is argued that
superradiance exists in the BTZ black hole (BTZbh) for the scalar
field. However, as we will show below, there is no superradiance
for vanishing boundary conditions at infinity.

The paper is organized as follows: in section 2, we give a
suggestive analysis of superradiance in BTZbh by following the
analysis of superradiance in Kerr given in \cite{bDW75}. With this
analysis no final conclusion about the existence of superradiance
in BTZbh for the real scalar field is reached. So in section 3, we
analyze superradiance for this system and vanishing boundary
conditions at infinity by using the exact solutions of the
Klein-Gordon operator in BTZbh. In this section we conclude that
superradiance does not exist for these boundary conditions.

\section{Superradiance?}

In order to achieve our goal we will use the known exact solutions
of the Klein-Gordon operator in the BTZbh, \cite{iIchiySat95} and
\cite{eKe99}. Before doing this we will analyze some properties of
the asymptotic solutions of the Klein-Gordon operator which shed
some light on the peculiarities of the problem under
consideration. We will do this by using some techniques borrowed
from \cite{bDW75}. The analysis of this section reflect how the
author came across with the problem of superradiance in BTZbh.
That is the reason for the title of this section. This section
sets the scene for asking if superradiance exists in BTZbh for the
real scalar field. The next section gives a definitive answer to
this question for one particular choice of boundary conditions at
infinity.

We shall assume that the scalar field $\varphi$ satisfies the
equation
\begin{equation}\label{E:36}
\left(\nabla_{\mu}\nabla^{\mu}-\xi R-m^{2}\right)\varphi=0,
\end{equation}
where $\xi$ is a coupling constant, $R$ is the Ricci scalar and
$m$ can be considered as the mass of the field. For the BTZ metric
\begin{equation}\label{E:2}
ds^{2}=-f^{2}dt^{2}+f^{-2}dr^{2}+r^{2}(d\phi+N^{\phi}dt)^{2},
\end{equation}
where
\begin{equation}\label{E:3}
f^{2}=\left(-M+\frac{r^{2}}{l^{2}}+\frac{J^{2}}{4r^{2}}\right)=\frac{\left(r^{2}-r_{+}^{2}\right)\left(r^{2}-r_{-}^{2}\right)}{l^{2}r^{2}}
\end{equation}
and
\begin{equation}\label{E:4}
N^{\phi}=-\frac{J}{2r^{2}}=-\frac{r_{+}r_{-}}{lr^{2}}
\end{equation}
where
\begin{equation}\label{E:5}
r_{\pm}^{2}=\frac{Ml^{2}}{2}\left(1\pm\left(1-\left(\frac{J}{Ml}\right)^{2}\right)^{1/2}\right),
\end{equation}
with $|J|\leq Ml$, the Ricci scalar is
$R=6\Lambda=-\frac{6}{l^{2}}$.

Hence the last equation can be written as
\begin{equation}\label{E:a37e}
\left(\nabla_{\mu}\nabla^{\mu}-\widetilde{m}^{2}\right)\varphi=0,
\end{equation}
where $\widetilde{m}^{2}=m^{2}-\frac{6\xi}{l^{2}}$. Here
$\widetilde{m}^{2}$ can be negative since we assume $m^{2}\geq 0$.
The operator $\nabla_{\mu}\nabla^{\mu}$ is given by
\begin{equation}\label{E:a38e}
\nabla_{\mu}\nabla^{\mu}\varphi=\frac{1}{\sqrt{|g|}}\partial_{\mu}\left(\sqrt{|g|}g^{\mu\nu}\partial_{\nu}\varphi\right),
\end{equation}
where $g=|g_{\mu\nu}|$. For the BTZ metric with $J\neq 0$ we have
$g=-r^{2}$, then $|g|=r^{2}$.

After a direct calculation, using (\ref{E:a38e}) in
(\ref{E:a37e}), it is obtained
\begin{equation}\label{E:a39e}
\frac{d^{2}R}{d{r^{\ast}}^{2}}+\{\left(\omega+nN^{\phi}\right)^{2}-f^{2}[\frac{n^{2}}{r^{2}}+\widetilde{m}^{2}+\frac{r^{1/2}}{2}\frac{d}{dr}\left(\frac{f^{2}}{r^{3/2}}\right)+\frac{f^{2}}{2r^{2}}]\}R=0,
\end{equation}
where it has been made the ansatz $\varphi(r,t,\phi)=e^{-i\omega
t}e^{in\phi}\frac{R(r)}{\sqrt{r}}$ and $r=r\left(r^{\ast}\right)$
with $\frac{dr^{\ast}}{dr}=f^{-2}$. This equation has been written
in \cite{sHyetal94}. However, there is a typo in the expression
given in this reference, the factor $r^{1/2}$ in the third term in
the square bracket is missing.

If (\ref{E:a39e}) is rewritten as
\begin{equation}\label{E:a42e}
\frac{d^{2}R}{d{r^{\ast}}^{2}}+V\left(r^{\ast}\right)R=0,
\end{equation}
where
\begin{equation}\label{E:43}
V\left(r^{\ast}\right)=\left(\omega+nN^{\phi}\right)^{2}-f^{2}[\frac{n^{2}}{r^{2}}+\widetilde{m}^{2}+\frac{r^{1/2}}{2}\frac{d}{dr}\left(\frac{f^{2}}{r^{3/2}}\right)+\frac{f^{2}}{2r^{2}}],
\end{equation}
then
\begin{equation}\label{E:a45e}
R_{1}\frac{dR_{2}}{dr^{\ast}}-R_{2}\frac{dR_{1}}{dr^{\ast}}=\textrm{const}
\end{equation}
where $R_{1}$ and $R_{2}$ are solutions of (\ref{E:a42e}). The
equation (\ref{E:a42e}) is valid for all the region
\textquotedblleft outside\textquotedblright the horizon where
$r^{\ast}$ goes from $-\infty$ to $0$. In analyzing superradiance
in Kerr metric, the equation (\ref{E:a45e}) is the starting point
\cite{bDW75}. The idea is to use this equation for two asymptotic
solutions of (\ref{E:a42e}) where in the asymptotic regions the
potential $V$ is finite. In the present case at the horizon, $N=0$
\begin{equation}\label{E:40}
\frac{d^{2}R}{d{r^{\ast}}^{2}}+\widetilde{\omega}^{2}R=0,
\end{equation}
where $\widetilde{\omega}=\omega+nN^{\phi}(r_{+})$. From the last
equation it follows that at the horizon the behavior of $R$ is
\begin{equation}\label{E:41}
R\propto e^{-i\widetilde{\omega}r^{\ast}}\qquad R\propto
e^{i\widetilde{\omega}r^{\ast}}.
\end{equation}
At first sight it seems that $V\left(r^{\ast}\right)$ goes to
$\infty$ when $r^{\ast} (r)$ goes to $0\,(\infty)$. This is
because at infinity the BTZbh is asymptotically AdS. So it seems
that we can not proceed further in the analysis by this method.
However from (\ref{E:a42e}) it follows that if $V$ were finite at
infinity then we would have
\begin{equation}\label{E:a46e}
R\propto e^{-i\omega' r^{\ast}}\qquad R\propto e^{i\omega'
r^{\ast}}
\end{equation}
for some $\omega'$. In this case we could analyze the
superradiance phenomenon in the same lines as in the Kerr case
using the techniques given in \cite{bDW75}. It turns out that $V$
tends to a constant value at infinity when
\begin{equation}\label{E:46ea}
\widetilde{m}^{2}+\frac{3}{4l^{2}}=0.
\end{equation}
Hence, in this case (\ref{E:a46e}) is true with
\begin{equation}\label{E:46eb}
\omega'=\sqrt{\omega^{2}-\frac{1}{l^{2}}\left(n^{2}+\frac{M}{4}\right)}.
\end{equation}
If at infinity there is a incident and a reflected wave
$R_{\infty}\propto e^{-i\omega' r^{\ast}}+Ae^{i\omega' r^{\ast}}$
with $A$ a complex constant, and at the horizon an incident wave
$R_{H}\propto Be^{-i\widetilde{\omega}r^{\ast}}$ with $B$ also a
complex constant, then after substituting this solution and its
complex conjugate in (\ref{E:a45e}) it is obtained
\begin{equation}\label{E:47}
1-|A|^{2}=\frac{\widetilde{\omega}}{\omega'}|B|^{2}.
\end{equation}
From this equation it follows that if $\widetilde{\omega}<0$ or
$\omega<n\Omega_{H}$ with $\Omega_{H}=-N^{\phi}$, the angular
velocity of the horizon, then the reflected wave has a bigger
amplitude than the incident one. At this stage it seems to exist
superradiance when (\ref{E:46ea}) is satisfied. However because of
(\ref{E:46eb}), it must be satisfied
\begin{equation}\label{E:47ea}
\omega>\frac{1}{l}\sqrt{n^{2}+\frac{M}{4}}.
\end{equation}
Also because $\omega<n\Omega_{H}$, it must be satisfied
\begin{equation}\label{E:47eb}
\omega<\frac{nJ}{Ml^{2}\left(1+\left(1-\left(\frac{J}{Ml}\right)^{2}\right)^{1/2}\right)},
\end{equation}
where we have used $\Omega_{H}=\frac{J}{2r_{+}^{2}}$. Because
$|J|\leq lM$ and $\omega>0$ both inequalities can not be satisfied
at the same time. Hence the fact that $\omega<n\Omega_{H}$ could
happen does not imply that superradiance exists. In the next
section we will show that it does not exist for vanishing boundary
conditions at infinity. We point out that these boundary
conditions are between the more natural ones since the BTZbh is
asymptotically AdS spacetime, and it has been shown
\cite{sjAvjIshdSto78} that a well defined quantization scheme can
be set up in AdS spacetime with these boundary conditions. Also
related with this issue is the fact that in four dimensions in the
Kerr-AdS black hole the existence of superradiance depends on the
boundary conditions at infinity \cite{eWins01}. So it is expected
that in the present case something analogous is happening.

\section{No superradiance in the BTZ black hole}

The discussion of this section follows closely the discussion in
\cite{iIchiySat95} and \cite{eKe99}, however in those works no
mention to superradiance is made.

If we assume harmonic dependence in $t$ and $\phi$, then the
operator (\ref{E:a38e}) reads
\begin{equation}\label{E:8}
\nabla_{\mu}\nabla^{\mu}\varphi=-\frac{1}{f^{2}r^{2}}\left(-\omega^{2}r^{2}+n^{2}\left(\frac{r^{2}}{l^{2}}-M\right)+n\omega
J\right)+\frac{1}{r}\partial_{r}\left(rf^{2}\partial_{r}\right)
\end{equation}
where we have made $\varphi(r,t,\phi)=e^{-i\omega
t}e^{in\phi}f_{n\omega}$. Hence the equation (\ref{E:a37e})
reduces to an equation in $r$ for $f_{n\omega}$
\begin{equation}\label{E:9}
\left[-\frac{1}{f^{2}r^{2}}\left(-\omega^{2}r^{2}+n^{2}\left(\frac{r^{2}}{l^{2}}-M\right)+n\omega
J\right)+\frac{1}{r}\frac{d}{dr}\left(rf^{2}\frac{d}{dr}\right)-\widetilde{m}^{2}\right]f_{n\omega}(r)=0.
\end{equation}

If we make $v=\frac{r^{2}}{l^{2}}$, then after some algebra we get
\begin{equation}\label{E:10}
\left(\frac{d^{2}}{dv^{2}}+\frac{\Delta'}{\Delta}\frac{d}{dv}+\frac{1}{4\Delta^{2}}\left(n\left(Mn-J\omega\right)-\widetilde{m}^{2}l^{2}\Delta-\left(n^{2}-\omega^{2}l^{2}\right)v\right)\right)f_{n\omega}(v)=0,
\end{equation}
where $\Delta=\left(v-v_{+}\right)\left(v-v_{-}\right)$ and
$'\equiv\frac{d}{dv}$. If now we let
\begin{equation}\label{E:10a}
f_{n\omega}=\left(v-v_{+}\right)^{\alpha}\left(v-v_{-}\right)^{\beta}g_{n\omega}
\end{equation}
we get
\begin{equation}\label{E:11}
u(1-u)g''_{n\omega}(u)+\left(c-\left(a+b+1\right)\right)g'_{n\omega}(u)-abg_{n\omega}(u)=0,
\end{equation}
where $u=\frac{v-v_{-}}{v_{+}-v_{-}}$,
$a=\alpha+\beta+\frac{1}{2}\left(1+\nu\right)$,
$b=\alpha+\beta+\frac{1}{2}\left(1-\nu\right)$, $c=2\beta+1$,
$\nu^{2}=1+\widetilde{m}^{2}l^{2}$,
$\alpha^{2}=-\frac{1}{4\left(v_{+}-v_{-}\right)^{2}}\left(r_{+}\omega-\frac{r_{-}n}{l}\right)^{2}$
and
$\beta^{2}=-\frac{1}{4\left(v_{+}-v_{-}\right)^{2}}\left(r_{-}\omega-\frac{r_{+}n}{l}\right)^{2}$.
The equation for $g_{n\omega}$ is the hypergeometric differential
equation, its solutions are well known. This equation has three
(regular) singular points at $0$, $1$, $\infty$ and two linear
independent solutions in a neighborhood of these points. Any of
these solutions can be analytically continued to another by using
the so-called linear transformation formulas, we will use this
property later. The solutions are divided in several cases
depending on the values of some combinations of the coefficients
$a$, $b$ and $c$. Let us consider the case when none of $c$,
$c-a-b$, $a-b$ is a integer.

The points $u=0,1,\infty$ correspond to the inner horizon, outer
horizon and infinity respectively. Because of the timelike
boundary of the BTZbh at infinity, we are interested in solutions
which allows us to have predictability, that is to say, we are
interested in situation where no new information coming from
infinity can enter to our problem. Let us consider the two
solutions at infinity. These solutions are given by
\begin{equation}\label{E:a12e}
g_{n\omega}=u^{-a}F(a,a-c+1;a-b+1;u^{-1})
\end{equation}
and
\begin{equation}\label{E:a13e}
g_{n\omega}=u^{-b}F(b,b-c+1;b-a+1;u^{-1}),
\end{equation}
where $F(a,b;c;z)$ is the hypergeometric function with
coefficients $a$, $b$ and $c$. If we write (\ref{E:10a}) as a
function of $u$ we have
\begin{equation}\label{E:a14e}
f_{n\omega}(u)=\left(v_{+}-v_{-}\right)^{\alpha+\beta}\left(u-1\right)^{\alpha}u^{\beta}g_{n\omega}(u)
\end{equation}
where $v_{\pm}=\frac{r_{\pm}^{2}}{l^{2}}$. Using (\ref{E:a14e}) in
(\ref{E:a12e}) and (\ref{E:a13e}) we have two functions at
infinity given by
\begin{equation}\label{E:15}
f_{n\omega}(u)=\left(v_{+}-v_{-}\right)^{\alpha+\beta}\left(u-1\right)^{\alpha}u^{\beta-a}F(a,a-c+1;a-b+1;u^{-1})
\end{equation}
and
\begin{equation}\label{E:a16e}
f_{n\omega}(u)=\left(v_{+}-v_{-}\right)^{\alpha+\beta}\left(u-1\right)^{\alpha}u^{\beta-b}F(b,b-c+1;b-a+1;u^{-1}).
\end{equation}
The last two equations can be approximated as
\begin{equation}\label{E:a17e}
f_{n\omega}(u)\sim\left(v_{+}-v_{-}\right)^{\alpha+\beta}u^{-h_{+}}F(a,a-c+1;a-b+1;u^{-1})
\end{equation}
and
\begin{equation}\label{E:18}
f_{n\omega}(u)\sim\left(v_{+}-v_{-}\right)^{\alpha+\beta}u^{-h_{-}}F(b,b-c+1;b-a+1;u^{-1}),
\end{equation}
where $h_{+}=\frac{1}{2}(1+\nu)$, $h_{-}=\frac{1}{2}(1-\nu)$ with
$\nu=\pm\sqrt{1+\widetilde{m}^{2}l^{2}}$. If we take the positive
square root then the first solution converges for any value of
$\nu$ and the second solution converges for $0\leq\nu<1$ and
diverges for $\nu\geq 1$. If we take the negative square root the
situation is inverted. Let us take the positive square root and
just the first solution. We can analytically continue (find the
expression) this solution to a neighborhood of $u=1$ using the
following linear relation \cite{mAbraiaSte65}
\begin{eqnarray}\label{E:19}
F(a,b;c;u)&=&\frac{\Gamma(c)\Gamma(a+b-c)}{\Gamma(a)\Gamma(b)}(1-u)^{c-a-b}u^{a-c}\times\nonumber\\
&\times&F(c-a,1-a;c-a-b+1;1-1/u)\\
          &+&\frac{\Gamma(c)\Gamma(c-a-b)}{\Gamma(c-a)\Gamma(c-b)}u^{-a}\times\nonumber\\
          &\times& F(a,a-c+1;a+b-c+1;1-1/u)\nonumber,
\end{eqnarray}
where $\Gamma(x)$ is the gamma function. By letting
$u\rightarrow\frac{1}{u}$, $a=a$, $b=a-c+1$ and $c=a-b+1$ in the
last equation we have
\begin{eqnarray}\label{E:20}
F(a,a-c+1;a-b+1;\frac{1}{u})&=&\frac{\Gamma(a-b+1)\Gamma(a+b-c)}{\Gamma(a)\Gamma(a-c+1)}\left(\frac{u-1}{u}\right)^{c-a-b}\times\nonumber\\
&\times & u^{1-b}F(1-b,1-a;c-a-b+1;1-u)\nonumber\\
&+&\frac{\Gamma(a-b+1)\Gamma(c-a-b)}{\Gamma(1-b)\Gamma(c-b)}u^{a}\times\\
&\times &F(a,b;a+b-c+1;1-u)\nonumber.
\end{eqnarray}
Inserting (\ref{E:20}) in (\ref{E:a17e}), close the outer horizon,
we have
\begin{eqnarray}\label{E:21}
f_{n\omega}&\sim&\frac{\Gamma(a-b+1)\Gamma(a+b-c)}{\Gamma(a)\Gamma(a-c+1)}\left(u-1\right)^{-\alpha}u^{-\beta}\times\nonumber\\
&\times & F(1-b,1-a;-2\alpha+1;1-u)\\
          &+&\frac{\Gamma(a-b+1)\Gamma(c-a-b)}{\Gamma(1-b)\Gamma(c-b)}\left(u-1\right)^{\alpha}u^{\beta}F(a,b;2\alpha+1;1-u)\nonumber.
\end{eqnarray}

The expression (\ref{E:21}) can be expressed as
\begin{eqnarray}\label{E:ae22}
f_{n\omega}&\sim
&\frac{\Gamma(1+\nu)\Gamma(2\alpha)}{\Gamma(\alpha+\beta+h_{+})\Gamma(\alpha-\beta+h_{+})}(u-1)^{-\alpha}u^{-\beta}\times\nonumber\\
&\times&F(-\alpha-\beta+h_{+},-\alpha-\beta+h_{-};-2\alpha+1;1-u)+\nonumber\\
&+&\frac{\Gamma(1+\nu)\Gamma(-2\alpha)}{\Gamma(-\alpha-\beta+h_{+})\Gamma(-\alpha+\beta+h_{+})}(u-1)^{\alpha}u^{\beta}\times\nonumber\\
&\times&F(\alpha+\beta+h_{+},\alpha+\beta+h_{-};2\alpha+1;1-u)
\end{eqnarray}
From this expression we can see that the two coefficients in both
terms are conjugate one of each other. Hence near $u=1$ we can
write the last expression as
\begin{equation}\label{E:ae23}
f_{n\omega}\sim
e^{i\theta}(u-1)^{\alpha}+e^{-i\theta}(u-1)^{-\alpha}
\end{equation}
where
$e^{2i\theta}=\frac{\Gamma(-\alpha-\beta+h_{+})\Gamma(-\alpha+\beta+h_{+})\Gamma(2\alpha)}{\Gamma(\alpha+\beta+h_{+})\Gamma(\alpha-\beta+h_{+})\Gamma(-2\alpha)}$.
We would like to write the last expression as a sum of two wave
modes. In order to do this we introduce another variable. First we
notice that
\begin{equation}\label{E:ae24}
\alpha=\pm\frac{i}{4\pi\textsf{T}}(\omega-\Omega n),
\end{equation}
where $\textsf{T}=\frac{r_{+}^{2}-r_{-}^{2}}{2\pi l^{2}r_{+}}$ and
$\Omega=\frac{r_{-}}{lr_{+}}$. We now define
$x=\frac{1}{4\pi\textsf{T}}\ln(u-1)$. With this definition the
equation (\ref{E:ae23}) becomes
\begin{equation}\label{E:ae25}
f_{n\omega}\sim e^{i\theta}e^{ix(\omega-\Omega
n)}+e^{-i\theta}e^{-ix(\omega-\Omega n)}.
\end{equation}
From here we conclude that the solution to the Klein-Gordon
operator near the outer horizon goes like
\begin{equation}\label{E:ae26}
\varphi\sim e^{-i\omega
t}e^{in\phi}\left(e^{i\theta}e^{ix(\omega-\Omega
n)}+e^{-i\theta}e^{-ix(\omega-\Omega n)}\right).
\end{equation}
From this expression we see that the mode near the outer horizon
is a superposition of an ingoing and an outgoing wave, both with
the same amplitude, hence cancelling each other. This is what we
expected since at infinity this mode vanishes, hence the
superradiance phenomenon does not appear.

It would be interesting to explore superradiance with other
fields, for example, the Dirac field. Also with the real scalar
field it would be interesting to study other boundary conditions
and see what happen.\vspace{0.5cm}

Acknowledgments: I thank my supervisor, Dr. Bernard S. Kay, his
guidance and several conversations during this work.

This work was carried out with the sponsorship of CONACYT Mexico
grant 302006.


\begin{thebibliography}{20}

    \bibitem{cMan87}
       C. Manogue, 
       Ann. Phys. \textbf{181}, 261 (1988).

    \bibitem{wUnruh74}
      W. Unruh, 
      Phys. Rev. D \textbf{10}, 3194 (1974).

    \bibitem{chLee77}
      C. H. Lee, 
      Phys. Lett. \textbf{68B}, 152 (1977).

    \bibitem{sCar95}
      S. Carlip, 
      Class. Quant. Grav. \textbf{12}, 2853 (1995). 

    \bibitem{iIchiySat95}
      I. Ichinose and Y. Satoh,
      Nucl. Phys. \textbf{B447}, 340 (1995).

    \bibitem{eKe99}
      E. Keski-Vakkuri,
      Phys. Rev. D \textbf{59}, 104001 (1999). 

    \bibitem{bDW75}
      B. DeWitt, 
      Phys. Rep. \text{19}, 295 (1975).

    \bibitem{sHyetal94}
      S. Hyun, G. H. Lee and J. H. Yee, 
      Phys. Lett. \textbf{322B}, 182 (1994).

    \bibitem{sjAvjIshdSto78}
      S. J. Avis, C. J. Isham and D. Storey, 
      Phys. Rev. D \textbf{18}, 3565 (1978).

    \bibitem{eWins01}
      E. Winstanley, 
      Phys. Rev. D \textbf{64}, 104010 (2001).

    \bibitem{mAbraiaSte65}
      M. Abramowitz and I. A. Stegun, \textit{Handbook of Mathematical
      Functions}
      (Dover, New York 1965)

\end{thebibliography}
\end{document}